\documentclass[final,numbers,sort&compress,3p,10pt]{elsarticle}

\usepackage{graphics}
\usepackage{float}
\usepackage{subfig}
\usepackage{graphicx}
\usepackage{booktabs}
\usepackage{multirow}
\usepackage{epsfig}
\usepackage{amssymb,algorithm,algorithmic}
\usepackage{mathtools}
\usepackage{xfrac}
\usepackage{bpchem}
\usepackage{amsfonts}
\usepackage{amsmath}
\usepackage{color,rotating}
\usepackage{ragged2e}
\usepackage{arydshln}
\usepackage{soul}
\usepackage{todonotes}
\usepackage{cancel}
\usepackage{lineno} 
\usepackage{tikz}
\usetikzlibrary{shapes,arrows}
\usetikzlibrary{fadings}
\usepackage{pgf}
\usepackage{xxcolor}
\usepackage{ctable}
\usepackage{hyperref,hyphenat }
\usepackage[nomarkers,notablist,nofiglist]{endfloat}

\newcommand{\cm}{\mathcal{M}}\newcommand{\fm}{\mathbf{M}}
\newcommand{\cd}{\mathcal{D}}

\newcommand{\eps}{\epsilon}
\newcommand{\etal}{\textit{et al.}}

\newcommand{\Bt}{\boldsymbol{\Theta}}
\newcommand{\indic}{\mathbb{I}_{\mathcal{N}_{\epsilon(y)}}(x)}
\newcommand{\indicprime}{\mathbb{I}_{\mathcal{N}_{\epsilon(y)}}(x')}

\journal{SIAM}
\begin{document}

\begin{frontmatter}

\title{Approximate Bayesian Computation by Subset Simulation} 

\author[UGR]{Manuel\,Chiachio\corref{cor1}}
\cortext[cor1]{Corresponding author. e-mail: mchiachio@ugr.es \\Tel: (+34)958240037 \,Fax: (+34)958249959.\\ E.T.S Ingenieros de Caminos, CC. y PP.\\ Campus de Fuentenueva s/n,~18071,~Granada,~Spain}
\author[CALTECH]{James L. Beck}
\author[UGR]{Juan Chiachio}
\author[UGR]{Guillermo Rus}

\address[UGR]{Dept. Structural Mechanics and Hydraulic Engineering, University of Granada.\\ Campus de Fuentenueva s/n, 18071 Granada, Spain.}
\address[CALTECH]{Division of Engineering and Applied Science, Mail Code 9-94, California Institute of Technology, Pasadena, CA 91125, USA. }
\begin{abstract}
A new~Approximate Bayesian Computation (ABC) algorithm for Bayesian updating of model parameters is proposed in this paper, which combines the ABC principles with the technique of \textit{Subset Simulation} for efficient rare\hyp event simulation, first developed in S.K.~Au and J.L.~Beck~\cite{SiuAu2001}.~It has been named ABC-SubSim.~The idea is to choose the nested decreasing sequence of regions in Subset Simulation as the regions that correspond to increasingly closer approximations of the actual data vector in observation space.~The efficiency of the algorithm is demonstrated in two examples that illustrate some of the challenges faced in real\hyp world applications of ABC.~We show that the proposed algorithm outperforms other recent sequential ABC algorithms in terms of computational efficiency while achieving the same, or better, measure of accuracy in the posterior distribution.~We also show that ABC-SubSim readily provides an estimate of the evidence~(marginal likelihood) for posterior model class assessment, as a by-product.
\end{abstract}


\end{frontmatter}


\section{Introduction}
\label{intro} The main goal of Bayesian statistics is to update a priori information about the parameter of interest~$\theta \in \Bt \subset \mathbb{R}^{d}$ for a parameterized model class $\cm$, based on the information contained in a set of data which we express as a vector~$y \in \cd  \subset \mathbb{R}^{\ell}$, where $\cd$ is the \textit{observation space}, the region in $\mathbb{R}^{\ell}$ of all possible observational outcomes according to the model class.~As a part of the model class $\cm$, we choose a prior probability density function (PDF) $p(\theta|\cm)$ over the parameter space and we also derive $p(y|\theta,\cm)$, the likelihood function of $\theta$, from the stochastic forward model $p(x|\theta,\cm)$ of the model class $\cm$ \cite{Beck2010}.~Bayes'~Theorem then yields the posterior PDF $p(\theta|y,\cm)$ of the  model specified by $\theta$ as follows:
\begin{equation}
p(\theta|y,\cm)=\frac{p(\theta|\cm)p(y|\theta,\cm)}{\int_{\Bt}p(\theta|\cm)p(y|\theta,\cm)d\theta }\propto p(\theta|\cm)p(y|\theta,\cm)
\label{eq:bayestheor}
\end{equation}
\noindent However, evaluation of the normalizing integral in the denominator is usually intractable except in some special cases.~Also, there are situations where Bayesian analysis is conducted with a likelihood function that is not completely known or it is difficult to obtain, perhaps because it requires the evaluation of an intractable multi\hyp dimensional integral over a latent vector, such as in hidden Markov models or dynamic state-space models, or because the normalization in the likelihood over the observation space $\cd$ involves an intractable integral parameterized by $\theta$ \cite{marin2011}.~Approximate Bayesian Computation (ABC) algorithms were conceived with the aim of evaluating the posterior density in those cases where the likelihood function is intractable \cite{tavare1997,pritchard1999}, although it also avoids the problem of the intractable integral in Equation \ref{eq:bayestheor}.~In the literature, these classes of algorithms are also called \textit{likelihood\hyp free computation algorithms}, which refers to their main aim of circumventing the explicit evaluation of the likelihood by using a simulation-based approach.~In this introductory section, we briefly summarize the body of ABC literature with a brief description of the main concepts and algorithms that we will need in the subsequent sections.

Let~$x \in\cd\subset\mathbb{R}^{\ell}$ denote a simulated dataset from~$p(\cdot|\theta,\cm)$, the forward model of model class $\cm$.~An ABC algorithm aims at evaluating the posterior~$p(\theta|y,\cm)\propto p(y|\theta,\cm)p(\theta|\cm)$~by applying Bayes' Theorem to the pair $(\theta,x)$:\begin{equation}
p(\theta,x|y) \propto p(y|x,\theta)p(x|\theta)p(\theta)
\label{eq:augmented}
\end{equation}
\noindent In the last equation, the conditioning on model class $\cm$ has been omitted for clarity, given that the theory is valid for any specific model class.~The function $p(y|x,\theta)$ gives higher weights for the posterior in those regions where $x$ is close to $y$.~The basic form of the algorithm to sample from the posterior given by Equation \ref{eq:augmented}, is a rejection algorithm that consists of generating jointly $\theta \sim p(\theta)$ and $x \sim p(x|\theta)$ and accepting them conditional on fulfilling the equality $x=y$.~Of course, obtaining sample $x=y$ is unlikely in most applications, and it is only feasible if $\cd$ consists of a finite set of values rather than a region in $\mathbb{R}^{\ell}$.~Hence two main approximations have been conceived in ABC theory to address this difficulty \cite{marjoram2003}:~a) replace the equality $x=y$ by the approximation $x \approx y$ and introduce a tolerance parameter $\eps$ that accounts for how close they are through some type of metric $\rho$; and~b) introduce a low-dimensional vector of summary statistics $\eta(\cdot)$ that permits a comparison of the closeness of $x$ and $y$ in a weak manner.~Through this approach, the posterior $p(\theta,x|y)$ in Equation \ref{eq:augmented} is approximated by $p_{\eps}(\theta,x|y)$, which assigns higher probability density to those values of $(\theta,x)\in \Bt \times \cd$ that satisfy the condition $\rho\big(\eta(x),\eta(y)\big) \leqslant \epsilon$.

\indent The~ standard version~ of ~the ABC algorithm takes~the approximate likelihood\footnote{In what follows, we use $P(\cdot)$ to denote probability whereas a PDF is expressed as $p(\cdot)$.}~$P_{\eps}(y|\theta,x)=P(x\in\mathcal{N}_{\eps}(y)|x)$, where $\mathcal{N}_{\eps}(y)=\big\{x \in \cd: \rho \big(\eta(x),\eta(y)\big)\leqslant \eps \big\}$.~From Bayes' Theorem, the approximate posterior~$p_{\eps}(\theta,x|y)=p\big(\theta,x|x\in \mathcal{N}_{\eps}(y)\big)$~is given by:\begin{equation}
p_{\eps}({\theta, x|y})\propto P(x \in \mathcal{N}_{\eps}(y)|x)p(x|\theta)p(\theta)
\label{eq:bayesth}
\end{equation}
\noindent where $P(x \in \mathcal{N}_{\eps}(y)|x)=\indic$, an indicator function for the set $\mathcal{N}_{\eps}(y)$ that assigns a value of 1 when~$\rho \big(\eta(x),\eta(y)\big) \leqslant \eps$ and~0~otherwise.~So the output of the ABC algorithm corresponds to samples from the joint probability density function:
\begin{equation}
\label{eq:genABCindi}
p_{\eps}({\theta, x|y})\propto p(x|\theta)p(\theta)\indic
\end{equation}
\noindent with ultimate interest typically being in the marginal approximate posterior:
\begin{equation}
\label{eq:marginalABC}
p_{\eps}({\theta|y})\propto p(\theta)\int_{\cd} p(x|\theta)\indic dx=P(x\in \mathcal{N}_{\eps}(y)|\theta)p(\theta)
\end{equation}
\noindent This integration need not be done explicitly since samples from this marginal PDF are obtained by taking the $\theta$ component of samples from the joint PDF in Equation \ref{eq:genABCindi} \cite{Robertbook04}.~Notice that the quality of the posterior approximation in Equations \ref{eq:genABCindi} and \ref{eq:marginalABC} depends on a suitable selection of the metric $\rho$, the tolerance parameter $\eps$ and, of special importance, the summary statistic $\eta(\cdot)$~\cite{fearnhead2012}.~A pseudocode to generate $N$ samples by the standard version of ABC algorithm is given in Algorithm \ref{algo:standABC}.
\begin{algorithm}
\caption{Standard ABC}
\label{algo:standABC}
\begin{algorithmic} 
\FOR {$t =1$ to $N$}
\REPEAT
\STATE 1.- Simulate $\theta'$ from $p(\theta)$
\STATE 2.- Generate $x' \sim p(x|\theta')$
 \UNTIL {$ \rho \big(\eta(x'),\eta(y)\big) \leqslant \eps$} \hfil \\
 Accept $(\theta', x')$
\ENDFOR
\end{algorithmic}
\end{algorithm}

\indent The choice of tolerance parameter $\eps$ is basically a matter of the amount of computational effort that the user wishes to expend but a possible guiding principle is described later at the end of~\S\ref{sec:controlpar}.~For $\eps$ sufficiently small ($\eps \rightarrow 0 $),  $\eta(x)\rightarrow\eta(y)$, and so all accepted samples corresponding to Equation \ref{eq:marginalABC} come from the closest approximation to the required posterior density $p(\theta|y)$, where the exactness is achieved when $\eta(\cdot)$ is a sufficient statistic.~This desirable fact is at the expense of a high computational effort (usually prohibitive) to get $\eta(x)=\eta(y)$ under the model $p(x|\theta)$.~On the contrary, as $\eps \rightarrow \infty$, all accepted observations come from the prior.~So, the choice of $\eps$ reflects a trade-off between computability and accuracy.

 \indent Several computational improvements have been proposed addressing this trade-off.~In those cases where the probability content of the posterior is concentrated over a small region in relation to a diffuse prior, the use of Markov Chain Monte Carlo methods (MCMC) \cite{Gilks96,neal1993,Gilks2005} has been demonstrated to be efficient \cite{marjoram2003}.~In fact, the use of a proposal PDF $q(\cdot|\cdot)$ over the parameter space allows a new parameter to be proposed based on a previous accepted one, targeting the stationary distribution $p_{\epsilon}(\theta|y)$.~The resulting algorithm, commonly called ABC-MCMC, is similar to the standard one (Algorithm \ref{algo:standABC}) with the main exception being the acceptance probability, which in this case is influenced by the MCMC acceptance probability as follows:
\begin{algorithm}
\caption{ABC-MCMC}
\label{algo:ABC-MCMC}
\begin{algorithmic} 
\STATE 1.- Initialize $(\theta^{(0)}, x^{(0)})$ from $p_{\eps}(\theta,x|y)$; e.g. use Algorithm \ref{algo:standABC}.
\FOR {$n =1$ to $N$}
\STATE 2.- Generate $\theta'\sim q(\theta|\theta^{(n-1)})$ and $x' \sim p(x|\theta')$.
\STATE 3.- Accept $(\theta',x')$ as $(\theta^{(n)},x^{(n)})$ with probability: \\
$\alpha=\text{min}\,\bigg\{1,\frac{P_{\eps}(y|x',\theta')p(\theta')q(\theta^{(n-1)}|\theta')}{P_{\eps}(y|x^{(n-1)},\theta^{(n-1)})p(\theta^{(n-1)})q(\theta'|\theta^{(n-1)})} \bigg\}$
\STATE \textbf{else} set $(\theta^{(n)},x^{(n)})=(\theta^{(n-1)},x^{(n-1)})$
\ENDFOR
\end{algorithmic}
\end{algorithm}

\noindent When $P_{\eps}(y|x,\theta)=\indic$, as in our case, the acceptance probability $\alpha$ is decomposed into the product of the MCMC acceptance probability and the indicator function:\begin{equation}
\label{eq:accrateindi}
\alpha=\text{min}\,\bigg\{1,\frac{p(\theta')q(\theta^{(n-1)}|\theta')}{p(\theta^{(n-1)})q(\theta'|\theta^{(n-1)})} \bigg\}\indicprime
\end{equation}
\noindent In this case, Step 3 is performed only if $x' \in \mathcal{N}_{\eps}(y)$.~The efficiency of this algorithm is improved with respect to the Standard ABC algorithm, but Equation~\ref{eq:accrateindi} clearly shows that the dependence upon~$\eps$~in the indicator function may lead to an inefficient algorithm for a good approximation of the true posterior.~In fact, given that $\alpha$ can only be non\hyp zero if the event $\rho \big(\eta(x^{'}),\eta(y)\big) \leqslant \eps$ occurs, the chain may persist in distributional tails for long periods of time if $\eps$  is sufficiently small, due to the acceptance probability being zero in Step 3 of Algorithm \ref{algo:ABC-MCMC}. 

\indent Some modifications to the ABC-MCMC scheme have been proposed \cite{bortot2007} that provide a moderate improvement in the simulation efficiency.~See \cite{sisson2010} for a complete tutorial about ABC-MCMC.~More recently, to overcome this drawback associated with ABC-MCMC, a branch of computational techniques have emerged to obtain high accuracy $(\eps \rightarrow 0)$ with a feasible computational burden by combining sequential sampling algorithms \cite{delmoral2006} adapted for ABC.~These techniques share a common principle of achieving computational efficiency by learning about intermediate target distributions determined by a decreasing sequence of tolerance levels $\eps_{1} >\eps_{2}>\ldots>\eps_{m}=\eps$, where the last is the desired tolerance $\eps$.~Table \ref{tab:SeqABC} lists the main contributions to the literature on this topic.~However, more research is needed to perform posterior simulations in a more efficient manner.
\begin{table}[h]
\centering
\begin{center}\footnotesize
\caption{\footnotesize\textit{Bibliography synoptic table about ABC with sequential algorithms.~Papers ordered by increasing date of publication.}}
 \renewcommand{\arraystretch}{1.3}
\begin{tabular}{lm{1.5cm}m{0.7cm}m{6cm}}
\toprule
\textbf{Paper} & \textbf{Algorithm} &\textbf{Year} & \hspace{4cm} \textbf{Notes} \\  \midrule
S.A.~Sisson \etal~\cite{sisson2007} & ABC-PRC & 2007 & Requires forward and a backward kernels to perturb the particles.~Uses a SMC sampler.~Induces bias.\\ 
T.~Toni \etal~\cite{toni2009} & ABC-SMC & 2009 & Does not require resampling steps in \cite{sisson2007}.~Based on sequential importance sampling.~Induces bias. \\ 
S.A.~Sisson \etal~\cite{sisson09} & ABC-PRC & 2009 &  This version incorporates an improved weight updating function.~Outperforms original in \cite{sisson2007}.  \\ 
M.A.~Beaumont \etal~\cite{beaumont2009}  & ABC-PMC & 2009 & Does not require a backward kernel as in the preceding works \cite{sisson2007,sisson09}.\\ 
M.~Baragatti \etal~\cite{baragatti2011} & ABC-PT & 2011 & Based on MCMC with exchange moves between chains.~Capacity to exit from distribution tails. \\
C.C.~Drovanti \etal~\cite{Drovandi2011} & Adaptive ABC-SMC & 2011 & Outperforms original in \cite{toni2009}.~Automatic determination of the tolerance sequence $\eps_{j}, j=\{1,\ldots,m\}$ and the proposal distribution of the MCMC kernel.\\
P.~Del Moral \etal~\cite{delmoral2012} & Adaptive ABC-SMC & 2012 & More efficient than ABC-SMC \cite{toni2009,Drovandi2011}.~Automatic determination of the tolerance sequence $\eps_{j}, j=\{1,\ldots,m\}$. \\
\bottomrule
\multicolumn{4}{l}{\footnotesize{PRC: Partial Rejection Control, SMC: Sequential Monte Carlo, PT: Parallel Tempering,}}\\
\multicolumn{4}{l}{\footnotesize{PMC: Population Monte Carlo.}}
\end{tabular}
\label{tab:SeqABC}
\end{center}
\end{table}

\indent~In this paper we introduce a new sequential algorithm,~called \textit{Approximate Bayesian Computation based on Subset Simulation} (ABC-SubSim), which combines the ABC principle with the technique of Subset Simulation \cite{SiuAu2001,SiuAu2003,SiuAu2007} to achieve computational efficiency in a sequential way.~The main idea is to link an ABC algorithm with a highly\hyp efficient rare\hyp event sampler that draws conditional samples from a nested sequence of subdomains defined in an adaptive and automatic manner.~ABC\hyp SubSim can utilize many of the improvements proposed in the recent ABC literature because of the fact that the algorithm is focused on the core simulation engine.

The paper is organized as follows.~Section \ref{sec:basisSS} reviews the theory underlying Subset Simulation and then the ABC-SubSim algorithm is introduced in Section~\ref{sec:SubSym}.~The efficiency of ABC\hyp SubSim is illustrated in Section~\ref{sec:examples} with two examples of dynamical models with synthetic data.~In Section~\ref{sec:discussion},~the performance of the algorithm is compared with some others in the recent ABC literature and the use of ABC-SubSim for posterior model class assessment is discussed.~Section \ref{sec:conclusions} provides concluding remarks.
\section{Subset Simulation method}
\label{sec:basisSS}
Subset Simulation is a simulation approach originally proposed to compute small failure probabilities encountered in reliability analysis of engineering systems~(e.g.~\cite{SiuAu2001,SiuAu2003,Ching2005}).~Strictly speaking, it is a method for efficiently generating conditional samples that correspond to specified levels of a performance function $g:\mathbb{R}^{d}\rightarrow \mathbb{R}$ in a progressive manner, converting a problem involving rare\hyp event simulation into a sequence of problems involving more frequent events. 

Let $F$ be the failure region in the $z$\hyp space, $z\in Z\subset \mathbb{R}^{d}$, corresponding to exceedance of the performance function above some specified threshold level $b$:
\begin{equation}
\label{eq:fairegion}
F=\{z\in Z:g(z)>b\}
\end{equation}
\noindent For simpler notation, we use~$P(F)\equiv P(z\in F)$.~Let us now assume that $F$ is defined as the intersection of $m$ regions $F=\bigcap_{j=1}^{m}F_{j}$, such that they are arranged as a nested sequence $F_{1}\supset F_{2}\ldots\supset F_{m-1} \supset F_{m}=F$, where $F_{j}=\{z\in Z:g(z)>b_{j}\}$, with $b_{j+1}>b_{j}$, such that $p(z|F_{j})\propto p(z)\mathbb{I}_{F_{j}}(z)$, $j=1,\ldots, m$.~The term $p(z)$ denotes the probability model for $z$.~When the event $F_{j}$ holds, then $\{F_{j-1},\ldots,F_{1}\}$ also hold, and hence $P(F_{j}|F_{j-1},\ldots,F_{1})=P(F_{j}|F_{j-1})$, so it follows that:\begin{equation}
P(F)=P\Big(\bigcap_{j=1}^{m}F_{j}\Big)=P(F_{1})\prod_{j=2}^{m}P(F_{j}|F_{j-1})
\label{eq:totevnt}
\end{equation}
\noindent where $P(F_{j}|F_{j-1})\equiv P(z \in F_{j}|z \in F_{j-1})$, is the conditional failure probability at the $(j-1)^{th}$ conditional level.~Notice that although the probability $P(F)$ can be relatively small, by choosing the intermediate regions appropriately, the conditional probabilities involved in Equation~\ref{eq:totevnt} can be made large, thus avoiding simulation of rare events.

In the last equation, apart from $P(F_{1})$, the remaining factors cannot be efficiently estimated by the standard Monte Carlo method (MC) because of the conditional sampling involved, especially at higher intermediate levels.~Therefore, in Subset Simulation, only the first probability $P(F_{1})$ is estimated by MC:
\begin{equation}
\label{eq:MC}
P(F_{1})\approx \bar{P}_{1}=\frac{1}{N}\sum_{n=1}^{N}\mathbb{I}_{F_{1}}(z_{0}^{(n)})\,, \,\,\,\,\,\,\,\,\,\, z^{(n)}_{0} \overset{\text{i.i.d.}}{\sim} p(z_{0})
\end{equation}
\noindent When $j\geqslant 2$, sampling from the PDF $p(z_{j-1}|F_{j-1})$ can be achieved by using MCMC at the expense of generating $N$ dependent samples, giving:
\begin{equation}
\label{eq:condfailestimtr}
P(F_{j}|F_{j-1})\approx\bar{P}_{j}=\frac{1}{N}\sum_{n=1}^{N}\mathbb{I}_{F_{j}}(z^{(n)}_{j-1})\,,\,\,\,\,\,\,\,\,\,\, z^{(n)}_{j-1}\sim p(z_{j-1}|F_{j-1})\end{equation}
\noindent where $\mathbb{I}_{F_{j}}(z_{j-1}^{(n)})$ is the indicator function for the region $F_{j}, j=1,\ldots, m$, that assigns a value of 1 when $g(z_{j-1}^{(n)})>b_{j}$, and 0 otherwise. 

Observe that the Markov chain samples that are generated at the $(j-1)^{th}$ level which lie in $F_{j}$ are distributed as $p(z|F_{j})$ and thus, they provide ``seeds'' for simulating more samples according to $p(z|F_{j})$ by using MCMC sampling with no burn\hyp in required.~As described further below, $F_{j}$ is actually chosen adaptively based on the samples~$\{z_{j-1}^{(n)},n=1,\ldots,N\}$~from $p(z|F_{j-1})$ in such a way that there are exactly $NP_{0}$ of these seed samples in $F_{j}$~\big(so $\bar{P}_{j}=P_{0}$ in Equation \ref{eq:condfailestimtr}\big).~Then a further $(\sfrac{1}{P_{0}}-1)$ samples are generated from $p(z|F_{j})$ by MCMC starting at each seed, giving a total of $N$ samples in $F_{j}$.~Repeating this process, we can compute the conditional probabilities of the higher-conditional levels until the final region $F_{m}=F$ has been reached.

To draw samples from the target PDF $p(z|F_{j})$ using the Metropolis algorithm, a suitable proposal PDF must be chosen.~In the original version of Subset Simulation \cite{SiuAu2001}, a modified Metropolis algorithm (MMA) was proposed that works well even in very high dimensions (e.g.~$10^{3}$\hyp $10^{4}$), because the original algorithm fails in this case (essentially all candidate samples from the proposal PDF are rejected\hyp see the analysis in \cite{SiuAu2001}).~In MMA, a univariate proposal PDF is chosen for each component of the parameter vector and each component candidate is accepted or rejected separately, instead of drawing a full parameter vector candidate from a multi\hyp dimensional PDF as in the original algorithm.~Later in \cite{SiuAu2003}, grouping of the parameters was considered when constructing a proposal PDF to allow for the case where small groups of components in the parameter vector are highly correlated when conditioned on any $F_{j}$.~An appropriate choice for the proposal PDF for ABC-SubSim is introduced in the next section.

It is important to remark that in Subset Simulation, an inadequate choice of the $b_{j}$\hyp sequence may lead to the conditional probability $P(F_{j}|F_{j-1})$ being very small (if the difference $b_{j}-b_{j-1}$ is too large), which will lead to a rare\hyp event simulation problem.~If, on the contrary, the intermediate threshold values were chosen too close so that the conditional failure probabilities were very high, the algorithm would take a large total number of simulation levels $m$ (and hence large computational effort) to progress to the target region of interest, $F$.~A rational choice that strikes a balance between these two extremes is to choose the $b_{j}$\hyp sequence adaptively \cite{SiuAu2001}, so that the estimated conditional probabilities are equal to a fixed value $P_{0}$~(e.g. $P_{0}=0.2$).~For convenience, $P_{0}$ is chosen so that $NP_{0}$ and $1/P_{0}$ are positive integers.~For a specified value of $P_{0}$, the intermediate threshold value $b_{j}$ defining $F_{j}$ is obtained in an automated manner as the $\left[(1-P_{0})N\right]^{th}$ largest value among the values $g(z_{j-1}^{(n)}),\,n=1,\ldots,N$,~so that the sample estimate of $P(F_{j}|F_{j-1})$ in Equation~\ref{eq:condfailestimtr} is equal to $P_{0}$.

\section{Subset Simulation for ABC}
\label{sec:SubSym}
Here we exploit Subset Simulation as an efficient sampler for the inference of rare events by just specializing the Subset Simulation method described in \S\ref{sec:basisSS} to ABC.~To this end, let us define $z$ as $z=(\theta,x)\in Z=\Bt\times\cd \subset \mathbb{R}^{d+\ell}$, so  that $p(z)=p(x|\theta)p(\theta)$.~Let also $F_{j}$ in \S\ref{sec:basisSS} be replaced by a nested sequence of regions $D_{j},\,j=1\ldots,m$, in $Z$ defined by:
\begin{equation}
\label{eq:regions}
D_{j}=\Big\{ z\in Z: x \in \mathcal{N}_{\eps_{j}}(y) \Big\}\equiv\Big\{ (\theta,x): \rho \big(\eta(x),\eta(y)\big)\leqslant \eps_{j} \Big\}
\end{equation}
\noindent with $D_{j}\subset \Bt\times\cd$ and $\rho$ is a metric on the set $\{\eta(x): x\in \cd\}$.~The sequence of tolerances $\eps_{1},\eps_{2},\ldots,\eps_{m}$, with $\eps_{j+1}<\eps_{j}$, will be chosen adaptively as described in \S\ref{sec:basisSS}, where the number of levels $m$ is chosen so that $\eps_{m}\leqslant \eps$, a specified tolerance.

As stated by Equation \ref{eq:genABCindi}, an ABC algorithm aims at evaluating the sequence of intermediate posteriors $p(\theta,x|D_{j}), \,j=1,\ldots,m$, where by Bayes' Theorem:
\begin{equation}
\label{eq:bayesabc}
p(\theta,x|D_{j})=\frac{P(D_{j}|\theta,x)p(x|\theta)p(\theta)}{P(D_{j})}\propto \mathbb{I}_{D_{j}}(\theta,x)p(x|\theta)p(\theta)
\end{equation}
\noindent Here, $\mathbb{I}_{D_{j}}(\theta,x)$ is the indicator function for the set $D_{j}$.~Notice that when $\eps \rightarrow 0$, $D_{m}$ represents a small closed region in $Z$ and hence $P(D_{m})$ will be very small under the model $p(\theta,x)=p(x|\theta)p(\theta)$.~In this situation, using MCMC sampling directly is not efficient due to difficulties in initializing the chain and in achieving convergence to the stationary distribution, as was described in \S \ref{intro} for ABC-MCMC.~This is the point at which we exploit the efficiency of Subset Simulation for ABC, given that such a small probability $P(D_{m})$ is converted into a sequence of larger conditional probabilities, as stated in Equations \ref{eq:totevnt},~\ref{eq:MC} and \ref{eq:condfailestimtr}.

\subsection{The ABC\hyp SubSim algorithm}
Algorithm \ref{algo:pseudoABCSUBSIM} provides a pseudocode implementation of ABC\hyp SubSim that is intended to be sufficient for most situations.~The algorithm is implemented such that a maximum allowable number of simulation levels ($m$) is considered in case the specified $\eps$ is too small.~The choice of~$\eps$ is discussed at the end of~\S\ref{sec:controlpar}.
\begin{algorithm}
\caption{Pseudocode implementation for ABC\hyp SubSim}
\label{algo:pseudoABCSUBSIM} 
\begin{algorithmic}
\STATE \underline{\textbf{\footnotesize{Inputs:}}}
\STATE $ P_{0} \in  [0,1]$ \COMMENT {gives percentile selection, chosen so $ NP_{0},\sfrac{1}{P_{0}} \in \mathbb{Z^{+}}$;~$P_{0}=0.2$ is recommended}. 
\STATE $N,$ \COMMENT {number of samples per intermediate level};~$m,$ \COMMENT {maximum number of simulation levels allowed}
\STATE \underline{\textbf{\footnotesize{Algorithm:}}}
\STATE Sample $\left[\big(\theta_{0}^{(1)}, x_{0}^{(1)}\big),\ldots,\big(\theta_{0}^{(n)}, x_{0}^{(n)}\big),\ldots,\big(\theta_{0}^{(N)}, x_{0}^{(N)}\big)\right]$, where $(\theta,x) \sim p(\theta)p(x|\theta)$ 
\FOR {$j:1,\ldots,m$}
\FOR {$n:1,\ldots,N$}
\STATE  Evaluate $\rho_{j}^{(n)}=\rho\big(\eta(x_{j-1}^{(n)}),\eta(y)\big)$  
\ENDFOR 
\STATE Renumber $\left[\big(\theta_{j-1}^{(n)},x_{j-1}^{(n)}\big), n:1,\ldots,N\right]$ so that $\rho_{j}^{(1)}\leqslant \rho_{j}^{(2)}\leqslant \ldots \rho_{j}^{(N)}$ 
\STATE Fix $\eps_{j}=\frac{1}{2} \left(\rho_{j}^{(NP_{0})}+\rho_{j}^{(NP_{0}+1)}\right)$
\FOR {$k=1,\ldots,NP_{0}$}
\STATE Select as a seed $\big(\theta^{(k),1}_{j},x^{(k),1}_{j}\big)=\big(\theta^{(k)}_{j-1},x^{(k)}_{j-1}\big)\sim p\big(\theta,x|(\theta,x) \in D_{j}\big)$
\STATE Run Modified Metropolis Algorithm \cite{SiuAu2001} to generate $1/P_{0}$ states of a Markov chain lying in $D_{j}$ (Eq. \ref{eq:regions}): $\left[\big(\theta_{j}^{(k),1},x_{j}^{(k),1}\big),\ldots,\big(\theta_{j}^{(k),1/P_{0}},x_{j}^{(k),1/P_{0}}\big)\right]$
 \ENDFOR
 \STATE Renumber $\left[(\theta_{j}^{(k),i},x_{j}^{(k),i}): k=1,\ldots,NP_{0};\, i=1,\ldots,1/P_{0} \right]$ as
 \STATE $\left[(\theta_{j }^{(1)},x_{j }^{(1)}),\ldots,(\theta_{j }^{(N)},x_{j }^{(N)})\right]$
\IF {$\eps_{j}\leqslant \eps$ } 
\STATE End algorithm \ENDIF
\ENDFOR
\end{algorithmic}
\end{algorithm}
\subsubsection{Choice of intermediate tolerance levels}
\label{sec:choiceinter}
 In Algorithm  \ref{algo:pseudoABCSUBSIM}, the~$\eps_{j}$ values are chosen adaptively as in Subset Simulation \cite{SiuAu2001}, so that the sample estimate $\bar{P}_{j}$ of $P(D_{j}|D_{j-1})$ satisfies $\bar{P}_{j}=P_{0}$.~By this way, the intermediate tolerance value $\eps_{j}$ can be simply obtained as the $100P_{0}$ percentile of the set of distances $\rho\big(\eta(x_{j-1}^{(n)}),\eta(y)\big), n=1,\ldots, N$,~arranged in increasing order.~Additionally, for convenience of implementation, we choose $P_{0}$ such that $NP_{0}$ and $\sfrac{1}{P_{0}}$ are integers, and so the size of the subset of samples generated in $D_{j-1}$  that lie in $D_{j}$ is known in advance and equal to $N P_{0}$.~These $NP_{0}$ samples in $D_{j}$ are used as seeds for $NP_{0}$ Markov chains of length $\sfrac{1}{P_{0}}$, where the new $(\sfrac{1}{P_{0}}-1)$ samples in $D_{j}$ in each chain are generated by MMA~\cite{SiuAu2001}.~Hence the total number of samples of $(\theta,x)$ lying in $D_{j}$ is $N$, but $NP_{0}$ of them were generated at the $(j-1)^{th}$ level.~Because of the way the seeds are chosen, ABC\hyp SubSim exhibits the benefits of \emph{perfect sampling} \cite{Konstia2012,Robertbook04}, which is an important feature to avoid wasting samples during a burn-in period, in contrast to ABC-MCMC.
\subsubsection {Choosing ABC\hyp SubSim control parameters}\label{sec:controlpar} The important control parameters to be chosen in Algorithm \ref{algo:pseudoABCSUBSIM} are $P_{0}$ and $\sigma_{j}^{2}$, the variance in the Gaussian proposal PDF in MMA at the $j^{th}$ level.~In this section we make recommendations for the choice of these control parameters.

In the literature, the optimal variance of a local proposal PDF for a MCMC sampler has been studied due to its significant impact on the speed of convergence of the algorithm \cite{Gelman1996,Roberts2001}.~ABC\hyp SubSim has the novelty of incorporating the Subset Simulation procedure in the ABC algorithm, so we use the same optimal adaptive scaling strategy as in Subset Simulation.~To avoid duplication of literature for this technique but conferring a sufficient conceptual framework, the method for the optimal choice of the~$\sigma_{j}^{2}$~is presented in a brief way.~The reader is referred to the recent work of \cite{Konstia2012}, where optimal scaling is addressed for Subset Simulation and a brief historical overview is also given for the topic. 
 
 \indent Suppose that the reason for wanting to generate posterior samples is that we wish to calculate the posterior expectation of a quantity of interest which is a function $h: \theta \in \Bt \rightarrow \mathbb{R}$.~We consider the estimate of its expectation with respect to the samples generated in each of the $j^{th}$ levels:\begin{equation}
\bar{h}_{j}=\mathbb{E}_{p_{\eps}(\theta|D_{j})}\left[h(\theta)\right]\approx \frac{1}{N}\sum_{n=1}^{N}h(\theta_{j}^{(n)})
\label{eq:estimh}
\end{equation}
\noindent where $\theta_{j}^{(n)}, n=1,\ldots,N$ are dependent samples drawn from $N_{c}$ Markov chains generated at the $j^{th}$ conditional level.~An expression for the variance of the estimator can be written as follows \cite{SiuAu2001}:
\begin{equation}
\label{eq:varsti}
\text{Var}(\bar{h}_{j})= \frac{R_{j}^{(0)}}{N}(1+\gamma_{j}),
\end{equation} 
  \noindent with 
\begin{equation}
 \label{eq:autocorfac}
\gamma_{j}=2\sum_{\tau=1}^{N_{s}-1}\left(\frac{N_{s}-\tau}{N_{s}}\right)\frac{R_{j}^{(\tau)}}{R_{j}^{(0)}}
\end{equation} 
\noindent In the last equation $N_{s}=\sfrac{1}{P_{0}}$ is the length of each of the Markov chains, which are considered probabilistically equivalent \cite{SiuAu2001}.~The term $R_{j}^{(\tau)}$ is the autocovariance of $h(\theta)$ at lag $\tau$, $R_{j}^{(\tau)}=\mathbb{E}\left[h(\theta_{j}^{(1)})h(\theta_{j}^{(\tau+1)})\right]-\bar{h}_{j}^{2}$, which can be estimated using the Markov chain samples $\big\{\theta_{j}^{(k),i}: k=1,\ldots,N_{c};\, i=1,\ldots,N_{s}\big\}$ as\footnote{It is assumed for simplicity in the analysis that the samples generated by the different $N_{c}$ chains are uncorrelated under the performance function $h$, although the samples are actually dependent because the seeds may be correlated.~See further details in \cite{SiuAu2001}, Section 6.2.}:
\begin{equation}
\label{eq:terms}
R_{j}^{(\tau)}\approx \tilde{R}_{j}^{(\tau)}=\left[\frac{1}{N-\tau N_{c}}\sum_{k=1}^{N_{c}}\sum_{i=1}^{N_{s}-\tau}h(\theta_{j}^{(k),i})h(\theta_{j}^{(k),\tau+i}) \right]-\bar{h}_{j}^{2}
\end{equation}

\noindent where $N_{c}=NP_{0}$, so that $N=N_{c}N_{s}$.

Given that the efficiency of the estimator $\bar{h}_{j}$ is reduced when $\gamma_{j}$ is high, the optimal proposal variance $\sigma_{j}^{2}$ for simulation level $j^{th}$ is chosen adaptively by minimizing $\gamma_{j}$.~This configuration typically gives an acceptance rate $\bar{\alpha}$ for each simulation level in the range of $0.2$\hyp$0.4$ \cite{Konstia2012}.~This is supported by the numerical experiments performed with the examples in the next section, which leads to our recommendation for ABC-SubSim:~\textit{Adaptively choose the variance~$\sigma_{j}^{2}$ of the $j^{th}$ intermediate level so that the monitored acceptance rate~$\bar{\alpha}\in\left[0.2,0.4\right]$ based on an initial chain sample of small length~(e.g. 10 states)}.

The choice of the conditional probability $P_{0}$ has a significant influence on the number of intermediate simulation levels required by the algorithm.~The higher $P_{0}$ is, the higher the number of simulation levels employed by the algorithm to reach the specified tolerance $\eps$, for a fixed number of model evaluations~($N$)~per simulation level.~This necessarily increases the computational cost of the algorithm.~At the same time, the smaller $P_{0}$ is, the lower the quality of the posterior approximation, that is, the larger the values of $\gamma_{j}$ in Equation \ref{eq:varsti}.~The choice of $P_{0}$ therefore requires a trade-off between computational efficiency and efficacy, in the sense of quality of the ABC posterior approximation.

To examine this fact, let us take a fixed total number of samples, i.e. $N_{T}=mN$, where $m$ is the number of levels required to reach the target tolerance value $\eps$, a tolerance for which $R_{m}^{(0)}\approx \text{Var}\left[h(\theta)\right]$.~The value of $m$ depends on the choice of $P_{0}$.~We can choose $P_{0}$ in an optimal way by minimizing the variance of the estimator $\bar{h}_{m}$ for the last simulation level:
\begin{equation}
\label{eq:varstPo}
\text{Var}(\bar{h}_{m})= \frac{R_{m}^{(0)}}{\sfrac{N_{T}}{m}}(1+\gamma_{m})\propto m(1+\gamma_{m})
\end{equation} 
\noindent Notice that $\gamma_{m}$ also depends upon $P_{0}$, although it is not explicitly denoted, as we will show later in \S\ref{sec:examples} (Figure \ref{fig:SensitMA}).~In the original presentation of Subset Simulation in \cite{SiuAu2001}, $P_{0}=0.1$ was recommended, and more recently in \cite{Konstia2012}, the range $0.1\leqslant P_{0} \leqslant 0.3$ was found to be near optimal after a rigorous sensitivity study of Subset Simulation, although the optimality there is related to the coefficient of variation of the failure probability estimate.~The value $P_{0}=0.2$ for ABC\hyp SubSim is also supported by the numerical experiments performed with the examples in the next section, where we minimize the variance in Equation~\ref{eq:varstPo}~as a function of $P_{0}$, which leads to the recommendation:~\textit{For ABC-SubSim, set the conditional probability $P_{0}=0.2$}.

Finally, it is important to remark that an appropriate final tolerance~$\eps$~may be difficult to specify a priori.~For these cases, one recommendation is to select~$\eps$~adaptively so that the posterior samples give a stable estimate $\bar{h}_{m}$~of~$\mathbb{E}_{p_{\eps}(\theta|D_{m})}\left[h(\theta)\right]$ (Equation \ref{eq:estimh}), i.e.~a further reduction in $\eps$ does not change $\bar{h}_{m}$ significantly.
\subsection{Evidence computation by means of ABC\hyp SubSim}
In a modeling framework, different model classes can be formulated and hypothesized to idealize the experimental system, and each of them can be used to solve the probabilistic inverse problem in Equation \ref{eq:bayestheor}.~If the modeler chooses a set of candidate model classes $\fm=\{\cm_{k},~k=1,\ldots,N_{M}\}$, Bayesian model class assessment is a rigorous procedure to rank each candidate model class based on their probabilities conditional on data~$y$~\cite{mackay1992,Beck2004}:
\begin{equation}
\label{eq:bayclass}
P(\cm_{k}|y,\fm)=\frac{p(y|\cm_{k})P(\cm_{k}|\fm)}{\sideset{}{}\sum_{i=1}^{N_{M}}p(y|\cm_{i})P(\cm_{i}|\fm)}
\end{equation}
\noindent where $P(\cm_{k}|\fm)$ is the prior probability of each $\cm_{k}$, that expresses the modeler's judgement on the initial relative plausibility of $\cm_{k}$ within $\fm$.~The factor $p(y|\cm_{k})$, which is called the \emph{evidence}~(or~\textit{marginal likelihood}) for the model class, expresses how likely the data $y$ are according to the model class.~The evidence $p(y|\cm_{k})$ is equal to the normalizing constant in establishing the posterior PDF in Equation \ref{eq:bayestheor} for the model class\footnote{The model parameter vector $\theta$ will, in general, be different for different model classes $\cm_{k.}$}: $p(y|\cm_{k})=\int_{\Bt}p(y|\theta,\cm_{k})p(\theta|\cm_{k})d\theta$.

When the likelihood is not available, the evidence $p(y|\cm_{k})$ is approximated using ABC by $P_{\eps}(y|\cm_{k})$, which depends upon $\eps$, the summary statistic $\eta(\cdot)$ as well as the chosen metric $\rho$ \cite{robert2011}.~In terms of the notation in Equation \ref{eq:regions}, the ABC evidence can be expressed as:
\begin{equation}
\label{eq:evidenceABC}
P_{\eps}(y|\cm_k)=P(D_{m}|\cm_{k})=\int_{\Bt}P(D_{m}|\theta,x,\cm_{k})p(x|\theta,\cm_{k})p(\theta|\cm_{k})d\theta dx
\end{equation}
The evaluation of the last integral is the computationally expensive step in Bayesian model selection, especially when $\eps\rightarrow 0$ \cite{marin2011}.~Observe that $P_{\eps}(y|\cm_k)$ in Equation \ref{eq:evidenceABC} is expressed as a mathematical expectation that can be readily estimated as follows:
\begin{equation}
\label{eq:MCABCMC}
P_{\eps}(y|\cm_k)\approx \frac{1}{N}\sum_{n=1}^{N}\mathbb{I}_{D_{m}}\big(\theta^{(n)},x^{(n)}\big)
\end{equation}
\noindent where $\big(\theta^{(n)},x^{(n)}\big)\sim p(x|\theta,\cm_{k})p(\theta|\cm_{k})$ are samples that can be drawn using the Standard ABC algorithm (Algorithm \ref{algo:standABC}), which in this setting is equivalent to the Standard Monte Carlo method for evaluating integrals.~The main drawback of this method arises when employing $\eps \rightarrow 0$, due to the well\hyp known inefficiency of the Standard ABC algorithm.~Moreover, the quality of the approximation in Equation \ref{eq:MCABCMC} may be poor in this situation unless a huge amount of samples are employed because otherwise the Monte Carlo estimator has a large variance.~Hence, several methods have emerged in the ABC literature to alleviate this difficulty, with the main drawback typically being the computational burden.~See \cite{didelot2011} for discussion of this topic.

ABC\hyp SubSim algorithm provides a straight-forward way to approximate the ABC evidence $P_{\eps}(y|\cm_{k})$ via the conditional probabilities involved in Subset Simulation:
\begin{equation}
\label{eq:ABCMCS}
P_{\eps}(y|\cm_k)=P(D_{m}|\cm_k)=P(D_{1})\prod_{j=2}^{m}P(D_{j}|D_{j-1})\approx P_{0}^{m}
\end{equation}
\noindent The last is an estimator for $P_{\eps}(y|\cm_k)$ which is asymptotically unbiased with bias $\mathcal{O}(\sfrac{1}{N})$.~See \cite{SiuAu2001,Konstia2012} for a detailed study of the quality of the estimators based on Subset Simulation where the approximation is studied in the context of the failure probability estimate~(but notice that Equations~\ref{eq:ABCMCS} and~\ref{eq:totevnt}~are essentially the same).~Of course, there are also approximation errors due to the ABC approximation that depend on the choice of~$\eps,~\eta(\cdot)$ and $\rho$ \cite{robert2011}.~Finally, once $P_{\eps}(y|\cm_{k})$ is calculated, it is substituted for $p(y|\cm_{k})$ in Equation \ref{eq:bayclass} to obtain $P_{\eps}(\cm_{k}|y,\fm)$, the ABC estimate of the model class posterior probability.~It is important to remark here that there are well\hyp known limitations of the ABC approach to the model selection problem, typically attributable to the absence of sensible summary statistics that work across model classes, among others \cite{didelot2011,robert2011}.~Our objective here is to demonstrate that calculation of the ABC evidence is a simple by\hyp product of ABC\hyp SubSim, as given in Equation \ref{eq:ABCMCS}.

\section{Illustrative examples}
\label{sec:examples}
In this section we illustrate the use of ABC\hyp SubSim with two examples:~1) a moving average process of order $d=2$, MA(2), previously considered in \cite{marin2011};~2)~a single degree\hyp of\hyp freedom (SDOF) linear oscillator subject to white noise excitation, which is an application to a state\hyp space model.~Both examples are input-output type problems, in which we adopt the notation $y=\left[y_{1},\ldots,y_{l},\ldots,y_{\ell}\right]$  for the measured system output sequence of length $\ell$.~The objective of these examples is to illustrate the ability of our algorithm to be able to sample from the ABC posterior for small values of $\epsilon$.~In the MA(2) example, we take for the metric the quadratic distance between the $d=2$ first autocovariances, as in \cite{marin2011}: \begin{equation}
\label{eq:MA2}
\rho\big(\eta(x),\eta(y)\big)=\sum_{q=1}^{d}(\tau_{y,q}-\tau_{x,q})^{2}
\end{equation}
\noindent In the last equation, the terms $\tau_{y,q}$ and $\tau_{x,q}$ are the autocovariances of $y$ and $x$, respectively, which are used as summary statistics.~They are obtained as $\tau_{y,q}=\sum_{k=q+1}^{\ell}y_{k}y_{k-q}$ and  $\tau_{x,q}=\sum_{k=q+1}^{\ell}x_{k}x_{k-q}$, respectively.~The Euclidean distance of $x$ from $y$ is considered as the metric for the oscillator example:
\begin{equation}
\label{eq:distance}
\rho(x,y)=\left[\sum_{l=1}^{\ell}(y_{l}-x_{l})^{2}\right]^{\sfrac{1}{2}}
\end{equation}
To evaluate the quality of the posterior, we study the variance of the mean estimator of a quantity of interest $h: \theta \in \Bt \rightarrow \mathbb{R}$, defined as follows (see \S\ref{sec:controlpar}):
\begin{equation}
\label{eq:perf}
h(\theta)=\sum_{i=1}^{d} \big(\theta_{i}\big)^{2}=\|\theta\|_{2}^{2},
\end{equation}

\subsection{Example 1:~Moving Average (MA) model}
\label{sec:MAex}
Consider a MA(2) stochastic process,~with $x_{l}, l=1,\ldots, \ell$, the stochastic variable defined by:\begin{equation}\label{eq:MA}
 x_{l}=e_{l}+\sum_{i=1}^{d}\theta_{i}e_{l-i}
\end{equation} 
\noindent with $d=2,\, \ell=100$ or $1000$.~Here $e$ is an i.i.d sequence of standard Gaussian distributions~$\mathcal{N}(0,1)$: $e=[e_{-d+1},\ldots,e_{0},e_{1}\ldots e_{l},\dots,e_{\ell}]$ and $x=[x_{1},\ldots,x_{l},\dots,x_{\ell}]$.~To avoid unnecessary difficulties, a standard identifiability condition is imposed on this model \cite{marin2011}, namely that the roots of the polynomial $\mathrm{D}(\xi)=1-\sum_{i=1}^{d}\theta_{i}\xi^{i}$ are outside the unit circle in the complex plane.~In our case of $d=2$, this condition is fulfilled when the region~$\Bt$~is defined as all~$(\theta_{1},\theta_{2})$ that satisfy: 
\begin{center}
$-2<\theta_{1}<2; \theta_{1}+\theta_{2}>-1; \theta_{1}-\theta_{2}<1 $
\end{center}
\noindent The prior is taken as a uniform distribution over $\Bt$.

Note that, in principle, this example does not need ABC methods as the likelihood is a multidimensional Gaussian with zero mean and a covariance matrix of order $\ell$ that depends on $(\theta_{1},\theta_{2})$, but its evaluation requires a considerable computational effort when $\ell$ is large \cite{marin2007book}.~This example was also used to illustrate the ABC method in \cite{marin2011} where it was found that the performance is rather poor if the metric is the one in Equation \ref{eq:distance} which uses the ``raw'' data but ABC gave satisfactory performance when the metric in Equation \ref{eq:MA2} was used.~For comparison with Figure 1 \cite{marin2011}, we also choose the latter here.

\begin{figure}[h]
\centering
\subfloat[\footnotesize\textit{$\ell=100$}]{\label{fig:ma1}\includegraphics[scale=1]{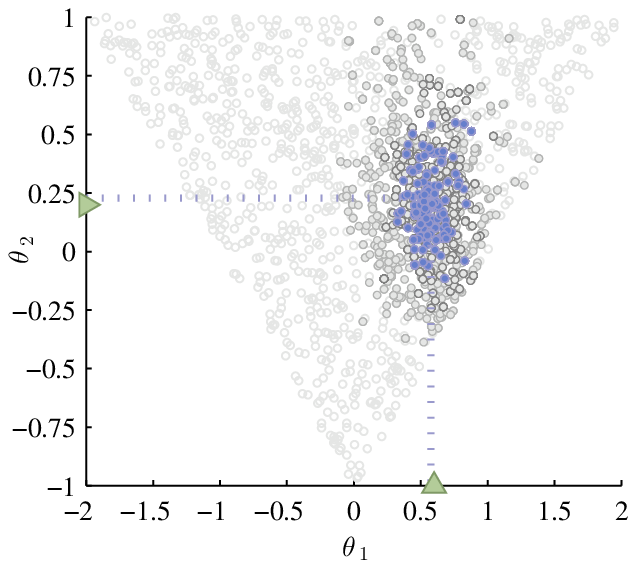}}
\subfloat[\footnotesize\textit{$\ell=1000$}]{\label{fig:ma2}\includegraphics[scale=1]{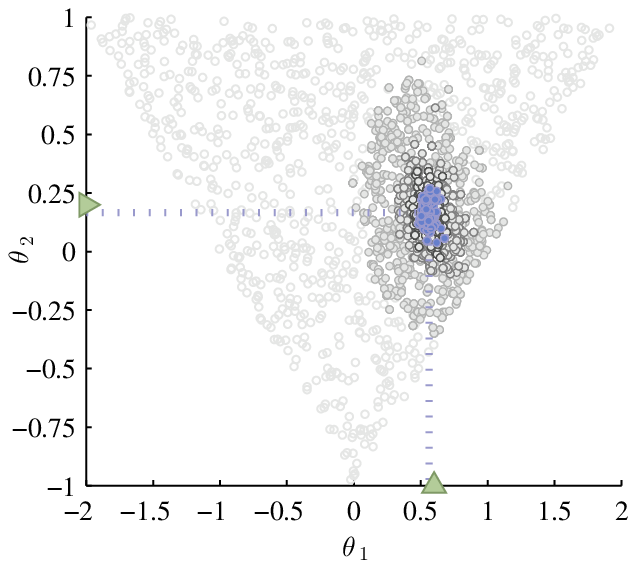}}
\caption{\footnotesize\textit{ABC\hyp SubSim output for the MA(2) model with (a) $\ell=100$ and (b) $\ell=1000$.~Each subplot presents samples (circles) in the model parameter space $\Bt$, where the latest final posterior samples are marked in dark blue circles.~The coordinates of the mean estimate of the latest posterior are represented in blue dotted line.~The green triangles are the coordinates of $\theta_{true}$.~To reveal the uncertainty reduction, the intermediate posterior samples are superimposed in increasing gray tones.~Gray rings correspond to prior samples.}}
\label{fig:ex1}
\end{figure}

We use synthetic data for $y$ by generating it from Equation \ref{eq:MA} considering $\theta_{true}=(0.6,0.2)$.~The chosen values of the control parameters for ABC\hyp SubSim are shown in Table \ref{tab:MAconf}.~The ABC\hyp SubSim results are presented in Figure \ref{fig:ex1}, which shows that the mean estimate of the ``approximate'' posterior samples at each level is close to $\theta_{true}$, for both $\ell=100$ and $\ell=1000$ cases.~Figure \ref{fig:ma1} shows the case $\ell=100$ which can be compared with Figure 1 in \cite{marin2011}.~In Figure \ref{fig:ma1}, a total of 3000 samples were used to generate 1000 samples to represent the posterior, whereas in \cite{marin2011}, 1,000,000 samples were used to generate 1000 approximate posterior samples using the standard ABC algorithm that we called Algorithm \ref{algo:standABC}.~The ABC\hyp SubSim posterior samples give a more compact set that is better aligned with the exact posterior contours given in Figure 1 of \cite{marin2011}.~Figure \ref{fig:ma2} shows that for the case $\ell=1000$, ABC\hyp SubSim used 4000 samples to generate 1000 samples representing the much more compact posterior that corresponds to ten times more data.
 \noindent 
\begin{table}[h]
\begin{center}\footnotesize
\caption{\footnotesize\textit{Parameter configuration of ABC\hyp SubSim algorithm for the MA(2) and SDOF linear oscillator examples.~The information shown in the first and second rows correspond to the MA(2) example with $\ell=100$ and $\ell=1000$, respectively.~The values shown from $4th$ to $7th$ column correspond to the optimal values for the proposal standard deviation per simulation level for both examples.}}
 \renewcommand{\arraystretch}{1.3}
\begin{tabular}{lccccccr}
\toprule
model&sample size& cond. probability& \multicolumn{4}{c}{proposal std. deviation}&  sim. levels \vspace{0.1cm}\\ 
&$(N)$&$(P_{0})$&$(\sigma_{1})$&$(\sigma_{2})$&$(\sigma_{3})$&$(\sigma_{4})$& $(m)$\\
\midrule
MA(2)~($\ell=100$)&$1000^{(*)}$ & $0.2$ &$0.4$&$0.2$&$0.1$& $--$  & $3$\\
MA(2)~($\ell=1000$)&$1000^{(*)}$ & $0.2$ &$0.4$&$0.2$&$0.1$& $0.04$  & $4$\\
Oscillator &$2000^{(*)}$ & $0.2$ &$0.35$&$0.1$&$0.05$& $0.001$  & $4$\\
\bottomrule 
\multicolumn{4}{l}{\footnotesize{(*): per simulation level}}
\end{tabular}
\label{tab:MAconf}
\end{center}
\end{table}

\begin{figure}[h]
\centering
\subfloat[\footnotesize$P_{0}=0.1$]{\label{fig:ex1sensp01}\includegraphics[scale=1]{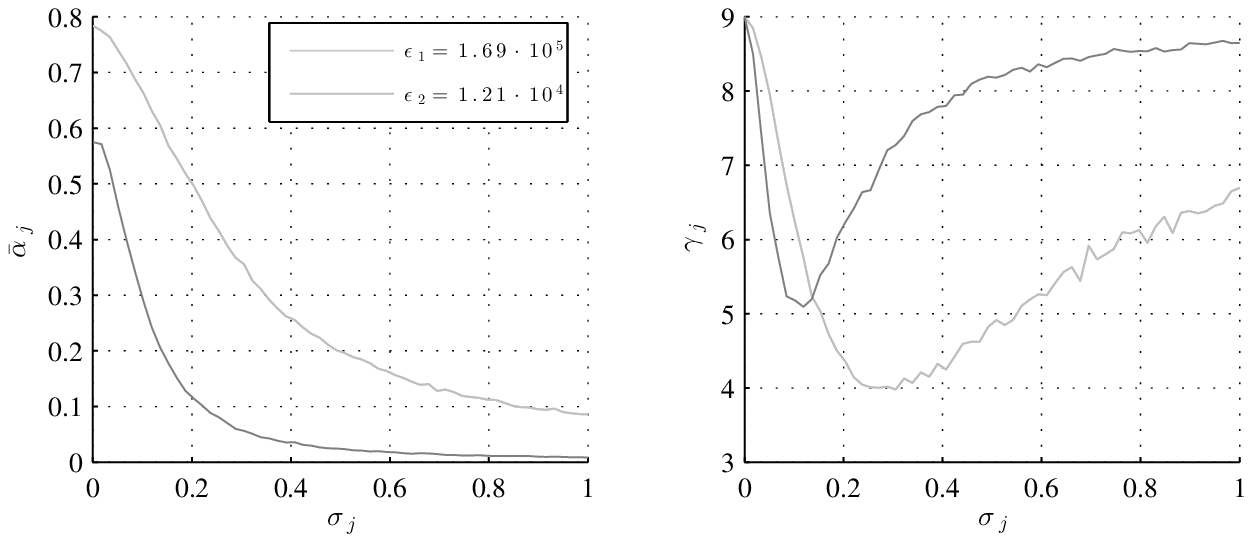}}\\
\subfloat[\footnotesize$P_{0}=0.2$]{\label{fig:ex1sensp02}\includegraphics[scale=1]{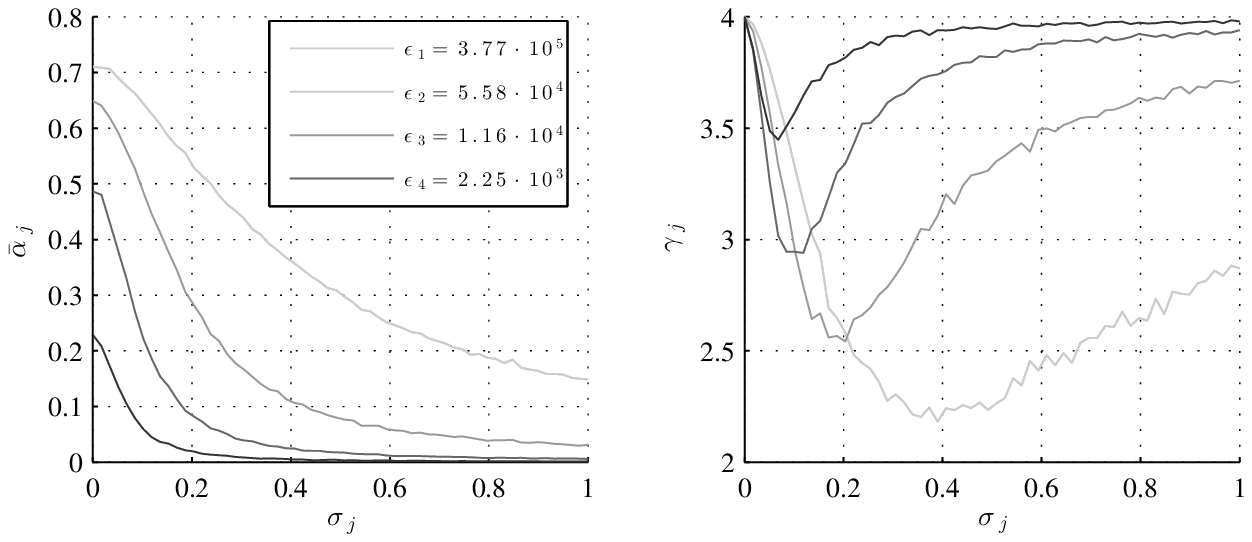}}\\
\subfloat[\footnotesize$P_{0}=0.5$]{\label{fig:ex1sensp05}\includegraphics[scale=1]{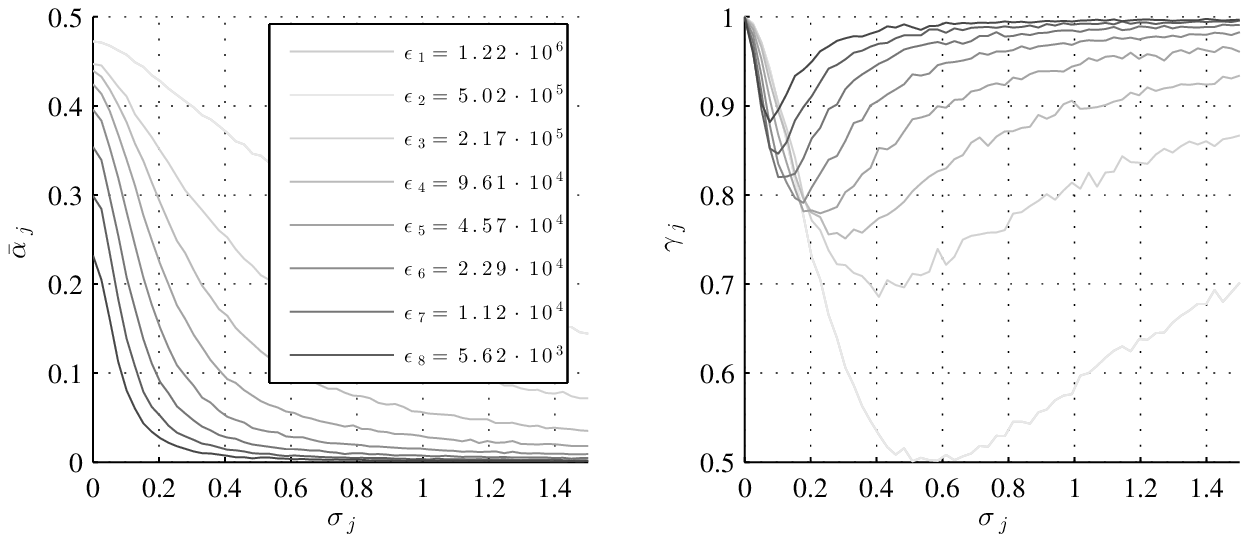}}
\caption{ \footnotesize\textit{Sensitivity study of the acceptance rate $\bar{\alpha}_{j}$ and autocorrelation factor $\gamma_{j}$ in relation to different values of the standard deviation $\sigma_{j}$ for the MA (2) model with $\ell=1000$ and for different values of $P_{0}=0.1\text{(a)},0.2\text{(b)},\text{and}~0.5\text{(c)}$.~$N=1000$ samples are employed per simulation level.~Darker curves correspond to higher simulation levels.~The tolerance values are indicated.~The numerical values of each plot are obtained considering the mean of 50 independent runs of the algorithm.}}
\label{fig:SensitMA}
\end{figure}

A preliminarily sensitivity study was done to corroborate the choice of the algorithm control parameters described in \S\ref{sec:controlpar} and the results are shown in Figure \ref{fig:SensitMA}.~As described in \S\ref{sec:controlpar}, the optimal value of $P_{0}$ is the one that minimizes $m(1+\gamma_{m})$ for fixed tolerance $\eps$.~As an exercise, we consider $\eps=1.12\cdot 10^{4}$ as the final tolerance.\footnote{It is unlikely that one or more values from the $\eps$\hyp sequence obtained using different $P_{0}$ values coincide exactly.~Hence, the nearest value to the final tolerance is consider for this exercise. }
The results in Figure \ref{fig:SensitMA} show that $P_{0}=0.2$ is optimal since then $m(1+\gamma_{m})=3(1+2.8)=11.4$; whereas for $P_{0}=0.5$ and $P_{0}=0.1$, it is $7(1+0.86)=13.1$ and $2(1+5.1)=12.2$, respectively.~These results are consistent with those for rare event simulation in \cite{Konstia2012}.~Observe also that the optimal variance $\sigma^{2}_{j}$ for the Gaussian proposal {\sc PDF} at the $j^{th}$ level that minimizes $\gamma_{j}$ occurs when the acceptance rate $\bar{\alpha}_{j}$ in {\sc MMA} lies in the range $0.2$\hyp$0.4$, which is also consistent with that found in \cite{Konstia2012} (except for the case of very low acceptance rate where the process is mostly controlled by the noise).
\subsection{Example 2:~Linear oscillator} 
\label{sec:SDOFexample}
Consider the case of a {\sc SDOF} oscillator subject to white noise excitation as follows:\begin{equation}
\label{eq:SDOF}
m\ddot{\xi}+c\dot{\xi}+k\xi=f(t)
\end{equation}
\noindent where  $\xi=\xi(t) \in \mathbb{R}~[m]$, $m~[Kg]$, $k~[\sfrac{N}{m}]$ and $c~[\sfrac{N\cdot s}{m}]$ are the displacement, mass, stiffness and damping coefficient, respectively.~To construct synthetic input, a discrete\hyp time history of input force $f~[N]$ modeled by Gaussian white noise with spectral intensity $S_{f}=0.0048\,[N^{2}\cdot s]$, is used.~The time step used to generate the input data is $0.01\,[s]$, which gives an actual value for the variance of the discrete input force $\sigma^{2}_{f}=3\,[N]$ \cite{hayes2009,Yuen2003}.

The probability model that gives the likelihood function of this example is Gaussian and so it can be written explicitly although its evaluation requires the computation of a high dimensional matrix inverse \cite{Kavenyuen2010}.~Repeated evaluations of the likelihood function for thousands of times in a simulation-based inference process is computationally prohibitive for large\hyp size datasets.~However it is easy to simulate datasets from this model after some trivial manipulations of Equation \ref{eq:SDOF} \cite{Kavenyuen2010}.~Therefore, this example is particularly suited for the use of ABC methods.

The mechanical system is assumed to have known mass $m=3~[Kg]$ and known input force giving the excitation.~For the state-space simulation, denote the \textit{state vector} by $s(t)=\left[\xi(t),\dot{\xi}(t)\right]^{T}$.~Equation~\ref{eq:SDOF} can be rewritten in state-space form as follows:\begin{equation}
\label{eq:ssSDOF}
\dot{s}(t)=A_{c}s(t)+B_{c} f(t)
\end{equation}
\noindent where $A_{c} \in \mathbb{R}^{2\times 2}$, $B_{c} \in \mathbb{R}^{2\times 1}$ are obtained as:
\begin{alignat}{2} 
A_{c}=\begin{pmatrix} 0 & 1  \\ -m^{-1}k & -m^{-1}c 
\end{pmatrix} & \qquad
B_{c}=\begin{pmatrix} 0 \\ m^{-1} 
\end{pmatrix}
\label{eq:matrixSDOF}
\end{alignat}
\noindent By approximating the excitation as constant within any interval, i.e.~$f(l\triangle t +\tau)=f(l\triangle t), \forall \tau \in \left[0,\triangle t\right)$, Equation\,\ref{eq:ssSDOF} can be discretized to a difference equation:~$\forall l\geqslant1$,\begin{equation}
s_{l} = As_{l-1}+Bf_{l-1}
\label{eq:discrSDOF}
\end{equation}
\noindent with $s_{l} \equiv s(l\triangle t)$, $f_{l}\equiv f(l\triangle t)$, $l=0,1,2,\ldots,\ell$,~and $A$ and $B$ are matrices given by:
\begin{subequations}
\begin{align} A &= e^{(A_{c}\triangle t)}\\
B&=A_{c}^{-1}\left(A-I_{2}\right)B_{c}
\label{eq:matrixSDOF2}
\end{align}
\end{subequations} 
\noindent where $I_{2}$ is the identity matrix of order 2.~The use of discrete\hyp time input and output data here is typical of the electronically-collected data available from modern instrumentation on mechanical or structural systems.

We adopt $\theta=\{k,c\}$ as unknown model parameters and denote by $y_{l}$ and $x_{l}$ the vectors consisting of the actual and predicted response measurements at each $\triangle t$.~Samples of $x_{l}$ for a given input force time history $\{f_{l}\}$ and $\theta$, can be readily generated by the underlying state-space model:
\begin{subequations}
\label{ref:trial}
\begin{align} s_{l}&=As_{l-1}+Bf_{l-1}+e_{l}\label{eq:modss}\\
x_{l}&=[1,0] s_{l}+e_{l}^{\prime}\label{eq:SSequations}
\end{align}
\end{subequations} 
\noindent where $e_{l}$ and $e_{l}^{\prime}$ are error terms to account for model prediction error and measurement noise, respectively.~Since in reality these errors would be unknown, we use the Principle of Maximum Information Entropy \cite{jaynes1957,jaynes2003,Beck2010} to choose $e_{l}$ and $e_{l}^{\prime}$ as i.i.d.~Gaussian variables, $e_{l} \sim \mathcal{N}(0,\sigma_{e}^{2}I_{2})$, $e_{l}^{\prime} \sim \mathcal{N}(0,\sigma_{e^{\prime}}^{2})$ and so they can be readily sampled.~For simplicity, we adopt $\sigma_{e}^{2}=10^{-2}$ and $\sigma_{e^{\prime}}^{2}=10^{-6}$, taking them as known.~We call $y=\{y_{1},\ldots,y_{l},\ldots, y_{\ell}\}$ the batch dataset collected during a total period of time $t=\ell\triangle t$, starting from known initial conditions $s_{0}=[0.01, 0.03]^{T}$ (units expressed in $[m]$ and $[m/s]$ respectively).~In this example, the noisy measurements $y_{l}$ are synthetically generated from Equation \ref{ref:trial} for the given input force history and for model parameters $\theta_{true}=\{k=4\pi,c=0.4\pi\}$.~We also adopt a sampling rate for the resulting output signal of $100 \left[\text{Hz}\right]$ ($\triangle t= 0.01 [s]$) during a sampling period of $t=3 [s]$, hence $\ell=300$.~We choose a uniform prior over the parameter space $\Bt$ defined by the region $0<\theta_{i}\leqslant3;\,i \in\{1,2\}$.~Table \ref{tab:MAconf} provides the information for the algorithm configuration.

\begin{figure}[h]
\centering
\subfloat{\label{fig:ex2}\includegraphics[scale=1]{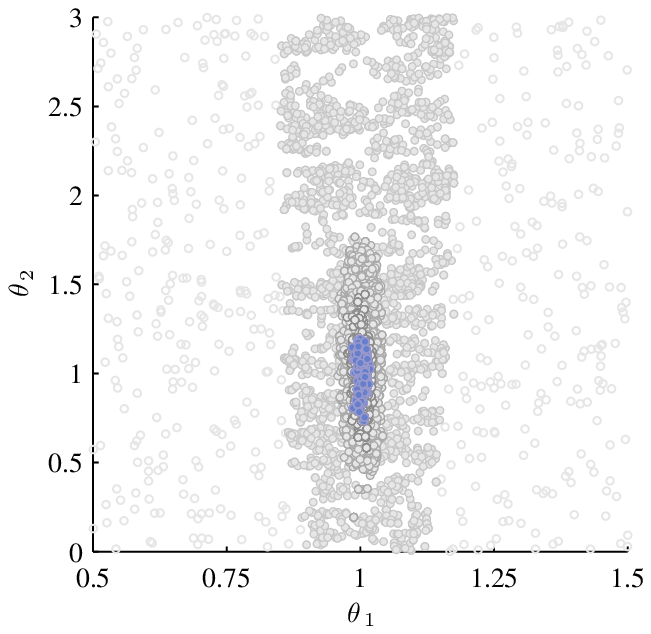}}
\subfloat{\label{fig:signal}\includegraphics[scale=1]{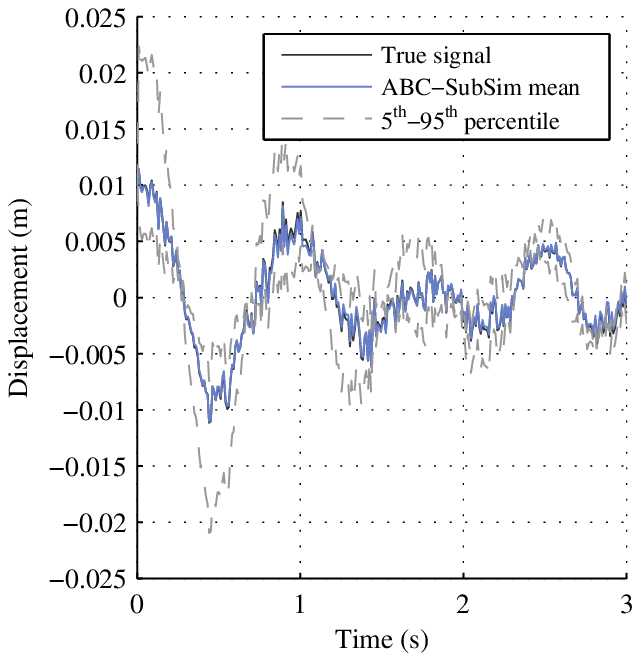}}
\caption{\footnotesize\textit{Results of the inference for the oscillator model for a duration of $t=3$ seconds.~Left: scatter plot of posterior samples of $\theta$ for intermediate levels and the final level (in blue).~The horizontal and vertical scale are normalized by a factor of $4\pi$ and $0.4\pi$, respectively.~Right: synthetic signal response of the oscillator, together with the mean estimate of the ABC\hyp SubSim approximation and two percentiles.}}
\label{fig:SDOF}
\end{figure}

The results shown in Figure \ref{fig:SDOF} are very satisfactory in the sense that ABC\hyp SubSim can reconstruct the true signal with high precision with only a moderate computational cost.~The posterior samples show that in Bayesian updating using noisy input-output data, the stiffness parameter $k=4\pi\theta_{1}$ is identified with much less uncertainty that the damping parameter $c=0.4\pi\theta_{2}$.~The normalized mean value over the set of posterior samples corresponding to the smallest value of $\eps$ is $\bar{\theta}=(1.00, 1.03)$, which is very close to the normalized true value $\theta_{true}=(1.0,1.0)$, (even if the exact likelihood was used, we would not expect $\bar{\theta}=\theta_{true}$ because of the noise in the synthetic data $y$).
\section{Discussion}
\label{sec:discussion}
\subsection{Comparison with recent sequential ABC algorithms}
In this section, ABC\hyp SubSim is compared with a selection of recent versions of sequential ABC algorithms:~ABC\hyp SMC \cite{delmoral2012},~ABC\hyp PMC \cite{beaumont2009} and ABC\hyp PT \cite{baragatti2011}, which are listed in Table \ref{tab:SeqABC}.~The same number of evaluations per simulation level are adopted for all algorithms, corresponding to 1000 and 2000 for the MA(2) and SDOF model, respectively.~We set the sequence of tolerance levels obtained by ABC\hyp SubSim using $P_{0}=0.5$ for the rest of the algorithms (see Table  \ref{tab:senstiresults}).~This was done because the recommended near-optimal value of $P_{0}=0.2$~(see~\S\ref{sec:controlpar}) for ABC-SubSim produced a sequence of $\eps$ values that decreased too quickly for ABC-PMC and ABC-SMC to work properly.~We note that this non-optimal choice of~$P_{0}$ for ABC-SubSim and the use of its~$\eps$-sequence provide considerable help for the competing algorithms.~The proposal PDFs are assumed to be Gaussian for all of the algorithms.~

\begin{figure}[h]
\centering
\subfloat{\label{fig:modevMA}\includegraphics[scale=1]{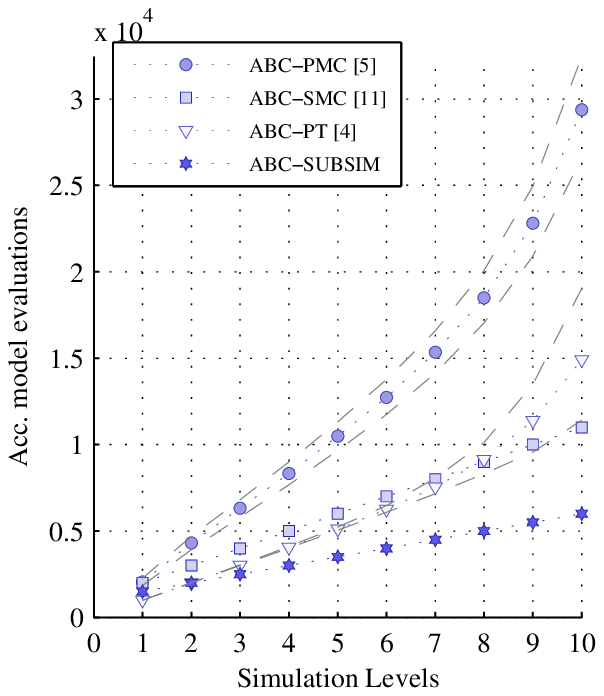}}\hspace{0.1cm}
\subfloat{\label{fig:varMA}\includegraphics[scale=1]{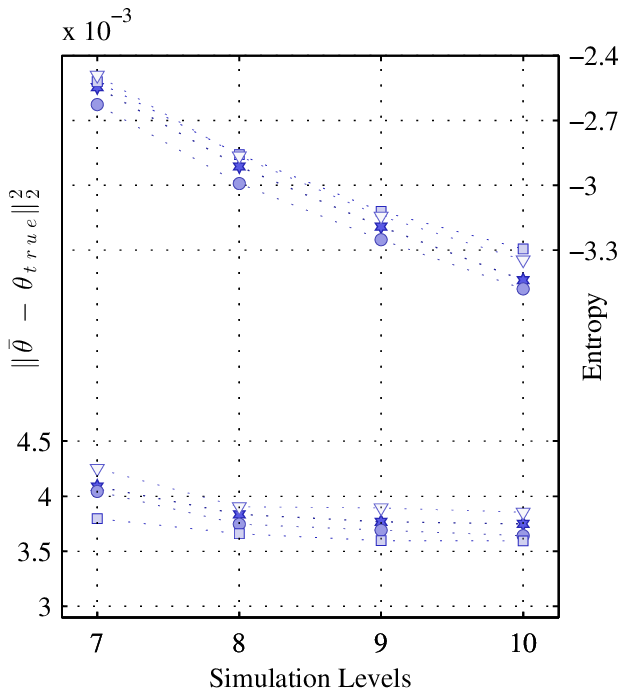}}\caption*{(a)~MA(2)}
\subfloat{\label{fig:modevSDOF}\includegraphics[scale=1]{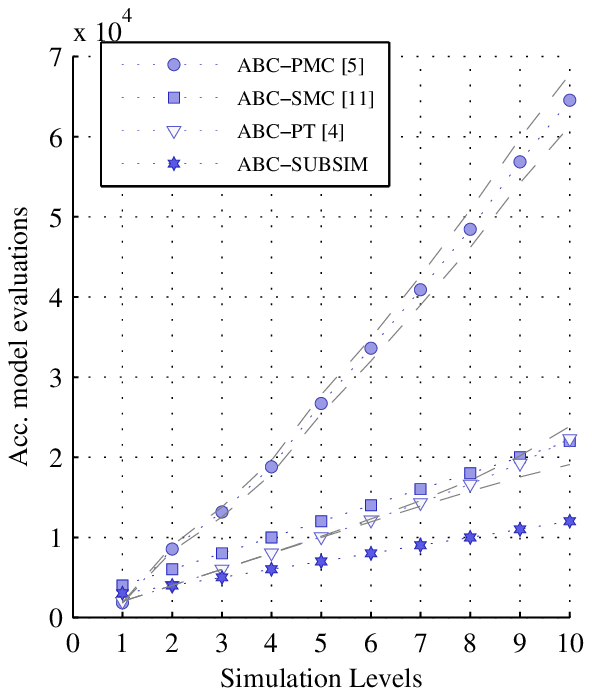}}\hspace{0.1cm}
\subfloat{\label{fig:varSDOF}\includegraphics[scale=1]{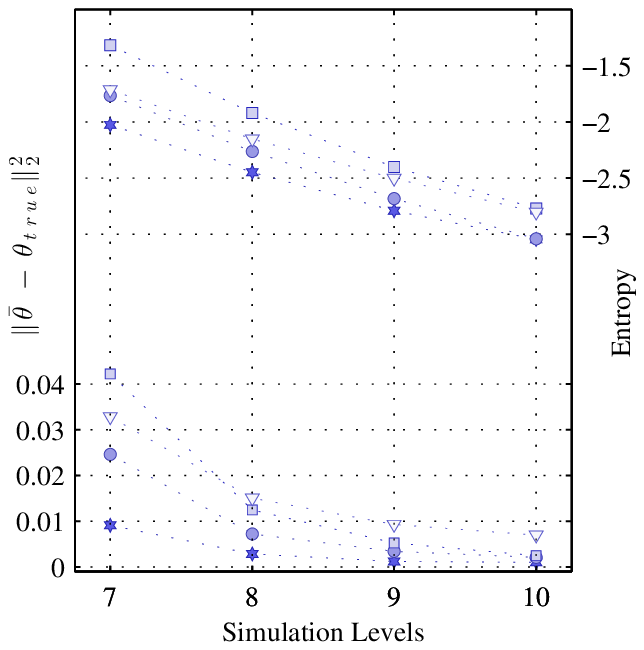}}\caption*{(b)~Oscillator}
\caption{\footnotesize\textit{Left: Accumulated model evaluations per simulation level for (a)~MA(2),~(b)~Oscillator.~Right: differential entropy (right\hyp side of the y\hyp label) of the intermediate posterior samples and mean quadratic error between $\bar{\theta}$ and $\theta_{true}$ (left\hyp side of the y\hyp label).~Both measures are evaluated for the last four intermediate simulation levels: $\eps_{j}, j=7,8,9,10$.~To be equivalent to ABC\hyp SubSim, we consider for the implementation of the ABC-SMC algorithm, a percentage of alive particles $\alpha=0.5$ and $M=1$ (see the details in \cite{delmoral2012}).}}
\label{fig:results}
\end{figure}

The results shown in Figure \ref{fig:results} are evaluated over the intermediate posterior samples for each simulation level and were obtained considering the mean of 100 independent runs of the algorithms, a large enough number of runs to ensure the convergence of the mean.~In this example, we focus on the number of model evaluations together with the quality of the posterior.~The left side of Figure \ref{fig:results} shows the accumulated amount of model evaluations employed by each of the competing algorithms.~Note that each algorithm requires the evaluation of auxiliary calculations, like those for the evaluation of particle weights, transition kernel steps, etc.~However, this cost is negligible because the vast proportion of computational time in ABC is spent on simulating the model repeatedly.~The number of model evaluations for ABC\hyp PMC and ABC\hyp PT is variable for each algorithm run, so in both cases we present the mean (labelled dotted lines) and a $95\%$ band (dashed lines).~In contrast, ABC\hyp SubSim and also ABC\hyp SMC make a fixed number of model evaluations at each simulation level.~Observe that the computational saving is markedly high when comparing with ABC\hyp PMC.

Regarding the quality of the posterior, we consider two measures: a) the sample mean of the quadratic error between $\bar{\theta}$ and $\theta_{true}$, i.e., $\|\bar{\theta}_{j}-\theta_{true}\|^{2}_{2}$, as an accuracy measure; and b) the differential entropy\footnote{This expression for the differential entropy is actually an upper-bound approximation to the actual differential entropy, where the exactness is achieved when the posterior PDF is Gaussian.} of the final posterior, by calculating $\sfrac{1}{2}\ln|(2\pi e)^{d}\det \left[cov(\theta_{j})\right]|$, as a measure quantifying the posterior uncertainty of the model parameters.~The results are shown on the right side of Figure \ref{fig:results}.~Only the last 4 simulation levels are presented for simplicity and clearness.

This comparison shows that ABC\hyp SubSim gives the same, or better, quality than the rest of the ABC algorithms to draw ABC posterior samples when $\eps$ is small enough, even though it used a smaller number of model evaluations.

\begin{table}[h]
\begin{center} \footnotesize
\caption{\footnotesize\textit{Set of tolerance values used for comparing the sequential ABC algorithms established using ABC\hyp SubSim with $P_{0}=0.5$.}}
\begin{tabular}{lcccccccccr}
\toprule
Model &$\eps_{1}$&$\eps_{2}$&$\eps_{3}$&$\eps_{4}$&$\eps_{5}$&$\eps_{6}$&$\eps_{7}$&$\eps_{8}$&$\eps_{9}$&$\eps_{10}$\\ \midrule
MA(2)~$(\times10^{4})$&$122$&$50.2$&$21.7$&$9.61$&$4.57$&$2.29$&$1.12$&$0.56$&$0.28$&$0.14$\\
Oscillator &$0.0117$&$0.0099$ &$0.0082$&$0.0054$&$0.0040$& $0.0030$ &$0.0024$&$0.0020$& $0.0018$& $0.0016$\\
\bottomrule 
\end{tabular}
\label{tab:senstiresults}
\end{center}
\end{table}

\subsection{Evidence calculation}
In this section we show how ABC\hyp SubSim algorithm can be applied to estimate the ABC evidence by taking advantage of the improvements in parameter space exploration introduced by Subset Simulation.~Table \ref{tab:evidence} shows the estimated values of the ABC evidence obtained with the ABC\hyp SubSim algorithm ($P_{0}=0.2$), which are computed using a total number of samples per simulation level $N$ equal to 1000 and 2000 for MA(2) and SDOF model, respectively.~For each value of $\eps_{j}$ chosen adaptively by ABC\hyp SubSim as described in \S\ref{sec:choiceinter}, we also calculate the ABC evidence using the approximation in Equation \ref{eq:MCABCMC} with $N=200,000$ samples per $\epsilon$ value for the Standard ABC algorithm~(a large enough amount of samples for the approximation in Equation \ref{eq:MCABCMC} to be sufficiently accurate).~It is seen in both examples that the results obtained by ABC-SubSim and Standard ABC agree well.

These results suggest that if the well\hyp known difficulties of the ABC model choice problem can be adequately resolved, high efficiency can be obtained by employing the ABC\hyp SubSim algorithm for the ABC evidence computation.

\begin{table}[h]
\begin{center}\footnotesize
\caption{\footnotesize\textit{Results of the estimation of the ABC evidence $P_{\eps_{j}}(\cd_{j}|\cm)$ for the MA(2) and oscillator examples when using 4 different tolerance values $\eps_{j}, j=1,\ldots,4$, which are produced by the ABC\hyp SubSim algorithm with $P_{0}=0.2$.~The Standard ABC algorithm employing 200,000 samples is also used to estimate $P_{\eps_{j}}(\cd_{j}|\cm)$ as in Equation \ref{eq:MCABCMC}}}
 \renewcommand{\arraystretch}{1.25}
 \scalebox{1}{

\begin{tabular}{lcccccr}
\toprule
    \multicolumn{3}{c}{ Example 1: MA(2)}&&\multicolumn{3}{c}{Example 2: Oscillator } \\
  \cmidrule(r){1-3} \cmidrule(r){5-7}  & SubSim  & Standard ABC & &&  SubSim  & Standard ABC  \\
\midrule
 $(\eps_{1}=3.77\cdot10^{5})$& 0.2 & 0.2070 & & ($\eps_{1}=0.0053)$ & 0.2 & 0.2038  \\
  $(\eps_{2}=5.58\cdot10^{4})$ &0.04 & 0.0412 && $(\eps_{2}=0.0023)$ & 0.04 &0.0397    \\
  $(\eps_{3}=1.16\cdot10^{4})$ & 0.008 & 0.0078 && $(\eps_{3}=0.0016)$ & 0.008 & 0.0079 \\
 $(\eps_{4}=2.25\cdot10^{3})$ & 0.0016 & 0.0017 && $(\eps_{4}=0.0014)$ & 0.0016 & 0.0016   \\
\bottomrule 
\end{tabular}}
\label{tab:evidence}
\end{center}
\end{table}

\section{Conclusions}
\label{sec:conclusions}
A new ABC algorithm based on Markov Chain Monte Carlo has been presented and discussed in this paper.~This algorithm  combines the principles of Approximate Bayesian Computation (ABC) with a highly\hyp efficient rare\hyp event sampler, Subset Simulation, which draws conditional samples from a nested sequence of subdomains defined in an adaptive and automatic manner.~We demonstrate the computational efficiency that can be gained with ABC\hyp SubSim by two different examples that illustrate some of the challenges in real-world applications of ABC.~The main conclusions of this work are:
\begin{itemize}
\item By its construction, ABC\hyp SubSim avoids the difficulties of ABC-MCMC algorithm in initializing the chain, as no burn-in is required.
\item In comparison with other recent sequential ABC algorithms, ABC\hyp SubSim requires a smaller number of model evaluations per simulation level to maintain the same quality of the posterior as the other algorithms. 
\item Together with ABC-SMC from \cite{delmoral2012}, ABC\hyp SubSim does not require the specification of a sequence of tolerance levels, which avoids tedious preliminary calibrations.
\item ABC\hyp SubSim allows a straightforward way to obtain an estimate of the ABC evidence used for model class assessment.
\end{itemize}

\section*{Acknowledgments}
The first two authors would like to thank the Education Ministry of Spain for the FPU grants~AP2009-4641, AP2009-2390 and also the California Institute of Technology (Caltech) which kindly hosted them during the course of this work.~The authors also acknowledge the Spanish Ministry of Economy for project DPI2010\hyp 17065 and also the European Union for the ``Programa Operativo FEDER de Andaluc\'ia 2007-2013'' for project GGI3000IDIB.

\newpage
\bibliographystyle{elsarticle-num-names}
\bibliography{ABC}   

\end{document}